\documentclass[pra,twocolumn,showpacs,preprintnumbers,amsmath,amssymb]{revtex4}
\usepackage{graphicx}
\usepackage{dcolumn}
\usepackage{bm}
\usepackage{latexsym,epsfig}
\usepackage{textcomp}

\begin{document}
\title{Mean field analysis of quantum phase transitions in a periodic optical superlattice}
\author{Arya Dhar}
\email{arya@iiap.res.in}
\author{Manpreet Singh}
\email{manpreet@iiap.res.in}
\affiliation{Indian Institute of Astrophysics, II Block, Koramangala, Bangalore-560 034, India}
\author{Ramesh V. Pai}
\email{rvpai@unigoa.ac.in} \affiliation{ Department of Physics, Goa
University, Taleigao Plateau, Goa 403 206, India. }
\author{B. P. Das}
\email{das@iiap.res.in} \affiliation{Indian Institute of
Astrophysics, II Block, Kormangala, Bangalore, 560 034, India.}

\date{\today}

\begin{abstract}
In this paper we analyze the various phases exhibited by a system
of ultracold bosons in a periodic optical superlattice using the mean field decoupling approximation.
 We investigate
for a wide range of commensurate and incommensurate densities. We
find the gapless superfluid phase, the gapped Mott insulator phase,
and gapped insulator phases with distinct density wave orders.

\end{abstract}



\maketitle

\section{Introduction}

Mean-field theory has proved to be a useful tool for the analysis of
the quantum phase transitions in the lattice
systems~\cite{sheshadri,jaksch}. The zero-temperature phase diagram
of the Bose-Hubbard model predicting the superfluid (SF) - Mott
insulator (MI) transition was first discussed by Fisher \textit{et
al} ~\cite{fisher}. Jaksch \textit{et al} ~\cite{jaksch} suggested
the possibility of such a transition in an optical lattice loaded
with ultra cold atoms and it was subsequently observed
experimentally by Greiner \emph{et al} in 2002~\cite{greiner}. There
are a number of reviews on this
topic~\cite{lewenstein,bloch,yukalov}. Several versions of the
mean-field theory have been used in the context of the ultracold
atoms; such as the Bogoliubov approximation~\cite{stoof}, the
Gutzwiller approach ~\cite{jaksch} and the mean-field decoupling
approximation~\cite{pai}. In the weak interaction limit, the
Bogoliubov approach is useful. However, it is not suitable for the
study of the SF-MI phase transition since it is valid only for weak
interactions. In the decoupling approximation, the Bose-Hubbard
Hamiltonian is decoupled into single-site Hamiltonians.
 The resulting mean-field equation can be
solved in two ways; either analytically using perturbation theory, or numerically
by diagonalizing the Hamiltonian matrix self consistently using a convenient basis. The Gutzwiller
mean-field approach has been used in several papers to study the
Bose-Hubbard model in quantum
lattices~\cite{jaksch,krauth,rokhsar}. In this paper, we have
applied the decoupling approximation to a d-dimensional periodic
optical superlattice with a periodicity of two
sites ~\cite{shuchen}.

A number of papers on the ultracold atoms in different types of
optical superlattices have been published in the past few
years~\cite{rousseau,roth,shuchen,schmitt,roux,arya}. Experiments on
this subject have been proposed and carried out in different
laboratories~\cite{piel, sebby,cheinet}. In this context, it is
desirable to understand the possible phases in different kinds of
optical superlattices. The main purpose of this study  is to
understand the phases in the d-dimensional optical superlattice with
a periodicity of two sites. For this purpose we use the mean-field
theory in the the decoupling approximation to convert the full
Hamiltonian into a sum of single cell Hamiltonians. Our findings for
ultracold atoms in the optical superlattice with a periodicity of 2
sites yields gapped insulators accompanied by different crystalline
orders in addition to the usual Mott insulator and superfluid
phases. These unusual insulating phases have been generically
referred to as the superlattice induced Mott insulators (SLMIs) in
the literature~\cite{arya}. Unlike the normal Mott insulator phase
where strong on-site interatomic interactions gives rise to the
gapped insulator, the SLMI phases arise due to the superlattice
potential. Depending upon the distribution of bosons within the unit
cell, there can be various types of SLMI phases. If the
configuration of the occupancy of bosons within the unit cell is
such that the alternate sites are occupied by one atom and the other
being empty, then such an insulator is called SLMI-I. The
configuration where the alternate sites are occupied by two bosons
and the other is empty, is called SLMI-II. If the configuration is
such that alternate sites are occupied by two bosons, and the other
by one, then it is called SLMI-III.

The rest of the paper is organized in the following
manner. In the next section, we describe  the application of the mean-field
decoupling approximation to an optical superlattice with a periodicity of two sites. In section
III we present our results. Our conclusions are given in the last
section; i.e. section IV.

\section{Mean-field Calculations for the Optical Superlattice}

The system of bosons in a general optical superlattice can be best
described by the Bose-Hubbard model as follows:-
\begin{equation}
 H=-t\sum_{\langle{i,j}\rangle}{(\hat{a}_{i}^{\dagger}\hat{a}_{j}+h.c)}+\frac{U}{2}\sum_{i}{\hat{n}_{i}(\hat{n}_{i}-1)}-{\sum_{i}\mu_i{\hat{n}_{i}}}
 \label{eq:ham}
\end{equation}

In the above equation, $\langle{i,j}\rangle$ denotes a pair of nearest
neighbor sites $i$ and $j$, $t$ denotes the  hopping amplitude
between adjacent sites, $U$ represents the on-site inter-atomic
interaction, $\hat{a}_{i}^{\dagger}$($\hat{a}_{i}$) is the creation
(annihilation) operator which creates (destroys) an atom at site $i$
and $\hat{n}_{i}=\hat{a}_{i}^{\dagger}\hat{a}_{i}$ is the number
operator, $\mu_i$ represents the on-site chemical potential. For an
optical lattice $\mu_i=\mu$ for all $i$. However, this is not true
for the optical superlattices and explicit dependence of $\mu_i$ on
the lattice site $i$ depends on the specifics of the superlattices.

The most important step in obtaining the
mean-field Hamiltonian is the decoupling of $\hat{a}_{i}^{\dag}\hat{a}_{j}$ into
single site operators. For this purpose, we make the following
approximation:-

\begin{displaymath}
a_{i}=\phi_{i}+\tilde{a}_{i};~~a_{i}^{\dagger}=\phi_{i}^{\ast}+\tilde{a}_{i}^{\dagger}
\end{displaymath}

Here $\phi_i=\langle{a_{i}}\rangle$, is the mean value and the
superfluid order parameter, and $\tilde{a}$ is the small fluctuation
over the mean value. We assume $\phi_i$ to be real~\cite{sheshadri},
hence, $\phi_i=\phi^{\ast}_i$ for all $i$. Substituting the above
approximation in the kinetic energy term of the Eq.~(\ref{eq:ham}),
we get
\begin{eqnarray}
& -& t\sum_{\langle{i,j}\rangle}{(a_{i}^{\dagger}a_{j}+h.c)} =  -t
\sum_{\langle{i,j}\rangle}{(\tilde{a}_{i}^{\dagger}\tilde{a}_{j}+\tilde{a}_{i}\tilde{a}_{j}^{\dagger})}\nonumber\\
&&- t\sum_{\langle{i,j}\rangle}(\tilde{a}_{i}^{\dagger}\phi_{j}
+\tilde{a}_{j}\phi_{i}+\tilde{a}_{i}\phi_{j}+\tilde{a}^\dagger_{j}\phi_{i}+2\phi_{i}\phi_{j})
\end{eqnarray}

We neglect the first term, which is second order in fluctuation. The
validity of such an approximation can be assumed when $t$ is small
compared to the the interaction $U$ and the superlattice potential
$\lambda$.
Defining $\bar{\phi}_i=\frac{1}{z}\sum_\delta\phi_{i+\delta}$,
$\delta$ being summed over $z=2d$ nearest neighbors with $d$ being
the dimension of the optical lattice,  we get the following
mean-field Hamiltonian,
\begin{eqnarray}
 H^{MF}&=&-tz \sum_{i}[{\bar\phi_i(\tilde{a}_{i}^{\dagger}+\tilde{a}_{i})+ \bar\phi_i \phi_i}]\nonumber \\
 && +\frac{U}{2} \sum_{i}{n_{i}(n_{i}-1)}-\sum_{i}\mu_i {n_{i}}
\end{eqnarray}

Substituting $\tilde{a}_{i}=a_{i}-\phi_i$ in the above equation, we
get

\begin{eqnarray}
 H^{MF}&=&-tz \sum_{i}[\bar\phi_i(a_{i}^{\dagger}+a_{i})- \bar\phi_i\phi_i]\nonumber \\
 &&+\frac{U}{2} \sum_{i}{n_{i}(n_{i}-1)}-\sum_{i}\mu_i{n_{i}}
\end{eqnarray}
which can be written as a  sum of single-site Hamiltonian, i.e.,
$H^{MF}=\sum_i H^{MF}_i$, where
\begin{equation}
 \frac{H_{i}^{MF}}{zt}=-\bar\phi_i(a_{i}^{\dagger}+a_{i})+\bar\phi_i\phi_i+\frac{\tilde{U}}{2}n_{i}(n_{i}-1)-\tilde{\mu}_i
 n_{i}.
\end{equation}
We have divided the single-site mean-field Hamiltonian by $zt$ to
make it and other parameters dimensionless, thus $\tilde{U}=U/zt$,
$\tilde{\mu}_i=\mu_i /zt$ are dimensionless on-site interaction and
chemical potential respectively.

For an optical lattice, all the sites are equal, thus $\mu_i=\mu$
and $\phi_i=\phi$ for all $i$. The Hamiltonian $H^{MF}_i$ can then
be diagonalized in the following manner. Assuming an initial value
for the superfluid order parameter $\phi$, the matrix elements of
the mean-field Hamiltonian is constructed in the number occupation
basis $|~n>$, where $n=0,1,2,\cdots,n_{max}$ , where $n_{max}$ is
the maximum number of bosons allowed per site whose value depends on
the on-site interaction $U$ and the chemical potential $\mu$.
Relatively small values of $n_{max}$ should suffice for large values
of $U$ and small values of $\mu$ and vice versa. Since we have taken
four different values of $U$ ranging from 2 to 15 and the density is
always less than 4, we have taken $n_{max}=10$ in our calculation.
The Hamiltonian matrix is then diagonalized to obtain the lowest
eigenstate which is used to obtain the new value for $\phi$. Using
this value for $\phi$, the calculation is repeated till $\phi$ is
converged.

We now extend the mean-field for the optical superlattice which is
formed by the superposition of two optical lattices with different
 wavelengths
 and a relative phase shift with respect to each other. This
 superlattice has a periodicity of two sites, thus each unit cell consists of
 $2^d$ sites with alternate sites have an energy shift $\lambda_i$.
 Such a system can be described by the Bose-Hubbard model (\ref{eq:ham})
 taking into account the relative energy shifts of the potential
 minima such that $\mu_i=\mu-\lambda_{i}$, $\lambda_i$ denotes the energy shift and is called the superlattice potential.

In this optical superlattice, the sites are not equal since
$\mu_i=\mu-\lambda_i$ is not same for all $i$. However, the
difference is restricted only within the unit cell and the whole
system is built using this unit cell. The unit cell in the
superlattice considered here consists of $2^d$ sites. Since all
dimensions are equivalent and the periodicity is two in our system,
we can work on any one such direction, and hence we denote
 the two neighboring sites by 1 and 2. The mean-field Hamiltonian for such a unit cell can be
 written as
\begin{eqnarray}
  H^{MF}_{uc} &=& -\bar\phi_2(a^\dagger_1+a_1)-\bar\phi_1(a^\dagger_2+a_2)+2\bar\phi_1\phi_2 \nonumber\\
   & &+
   \frac{\tilde{U}}{2}[n_1(n_1-1)+n_2(n_2-1)]\nonumber \\
   & & -\tilde\mu[n_1+n_2]+\tilde\lambda_1n_1+\tilde\lambda_2n_2
\end{eqnarray}

In order to diagonalize the above Hamiltonian, we express all the
operators including the Hamiltonian in the occupation number basis.
Then we take initial guess values of the superfluid order
parameters, $\phi_{1}$ and $\phi_{2}$. After diagonalising the
$H^{MF}_{uc}$ matrix using the standard Jacobi method, we find the
ground state energy, and the ground state wave-function. From the
relation $\phi_{i}=\langle{a_{i}}\rangle$, we calculate the
superfluid order parameters using the ground state wave-function. We
then substitute these new values of $\phi_{1}$ and $\phi_{2}$ in the
$H_{uc}$ and iterate the process, until the values of $\phi_{1}$ and
$\phi_{2}$ converge to $10^{-6}$. The different phases are then
analysed based on the values of these superfluid densities.
\begin{figure}[!t]
 \centering
\psfig{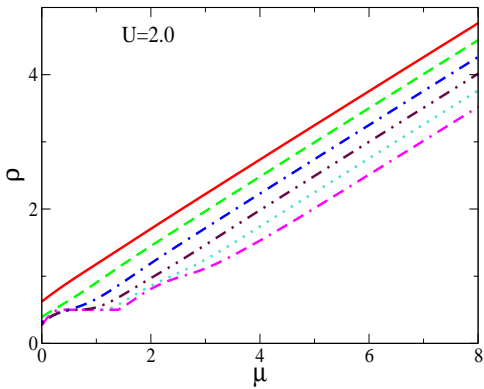}
\caption{(Color online) Variation of average density $\rho$ as a
function of the chemical potential $\mu$ for U=2, but for different
values of $\lambda$ starting from 0.5 (red solid curve) to 5.5 (magenta double dash dot
curve) at the intervals of 1.0.} \label{fig:fig1}
\end{figure}
\begin{figure}[!b]
 \centering
\psfig{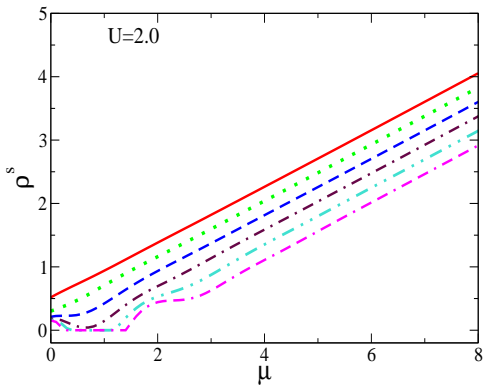}
\caption{(Color online) Variation of average superfluid density
$\rho^s$ as a function of $\mu$ for the same set of parameters as in
Fig.~\ref{fig:fig1}.} \label{fig:fig2}
\end{figure}
\section{Results}
Taking the superlattice potential for the two distinguished sites
within the unit cell  $\lambda_1=0$ and $\lambda_2=\lambda$, we
present our results for a wide range of $\lambda$, densities $\rho$
and four characteristic values of the on-site interaction $U$,
chosen to cover a substantial part of the phase diagram. In our analysis we have taken $t=1.0$, so all other quantities are expressed in units of $t$.

First we investigate the effect of the superlattice potential on the
superfluid phase. In the Figs.~\ref{fig:fig1} and \ref{fig:fig2}, we
plot, respectively, for $U=2$ the average density $\rho$ and the
superfluid density $\rho^s$ as a function of the chemical potential
$\mu$ for different values of $\lambda$ starting from $0.5$ to $5.5$
at an interval of $1.0$. Here $\rho$ and $\rho^s$ are,
respectively, the average density and the superfluid density of a
unit cell, i.e., $\rho=(\rho_1+\rho_2)/2$,
$\rho^s=(\phi^2_1+\phi^2_2)/2$. It is known that for
$U=2$, $\lambda=0$, the model~(\ref{eq:ham}) is always in the
superfluid phase~\cite{sheshadri} irrespective of the value of the
density $\rho$.

From the Figs.~\ref{fig:fig1} and \ref{fig:fig2}, we find that
density $\rho$ increases with increase in $\mu$ for finite, but
small values of $\lambda$. The superfluid density $\rho^s$ remains
finite for all densities which  imply that the system continues to
be in the superfluid phase as in the case of $\lambda=0$. However,
as
 $\lambda$ is increased further, say for example, $\lambda=4.5$,
the density develops a plateau at $\rho=1/2$ for a range of $\mu$
values. This is the signature of a finite gap in the energy spectrum
in this range of $\mu$ values and vanishing of compressibility
$\kappa=\frac{\partial \rho}{\partial \mu}$. The Fig.~\ref{fig:fig2}
suggest vanishing of superfluid density in the same range of $\mu$.
For all other densities, i.e., $\rho\ne 1/2$, including integer
densities, the superfluid density remains finite. These features
confirm that, for $U=2$, model~(\ref{eq:ham}) is in the superfluid
phase for all values of $\rho\ne 1/2$ for all values of $\lambda$.
However, for $\rho=1/2$ there is a superfluid to an insulator phase
transition as $\lambda$ is increased.
\begin{figure}[!t]
 \centering
\psfig{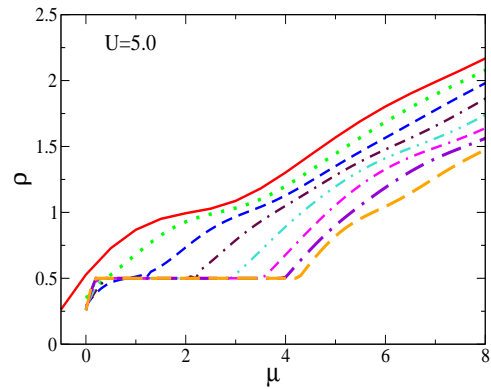}
\caption{(Color online) Variation of average density of a unit cell
$\rho$ as a function of the chemical potential $\mu$ for U=5, but
for different values of $\lambda$ starting from 0.2 (red solid curve) to
7.2 (orange large dashed curve) at intervals of 1.0. } \label{fig:fig3}
\end{figure}
\begin{figure}[!b]
 \centering
\psfig{file=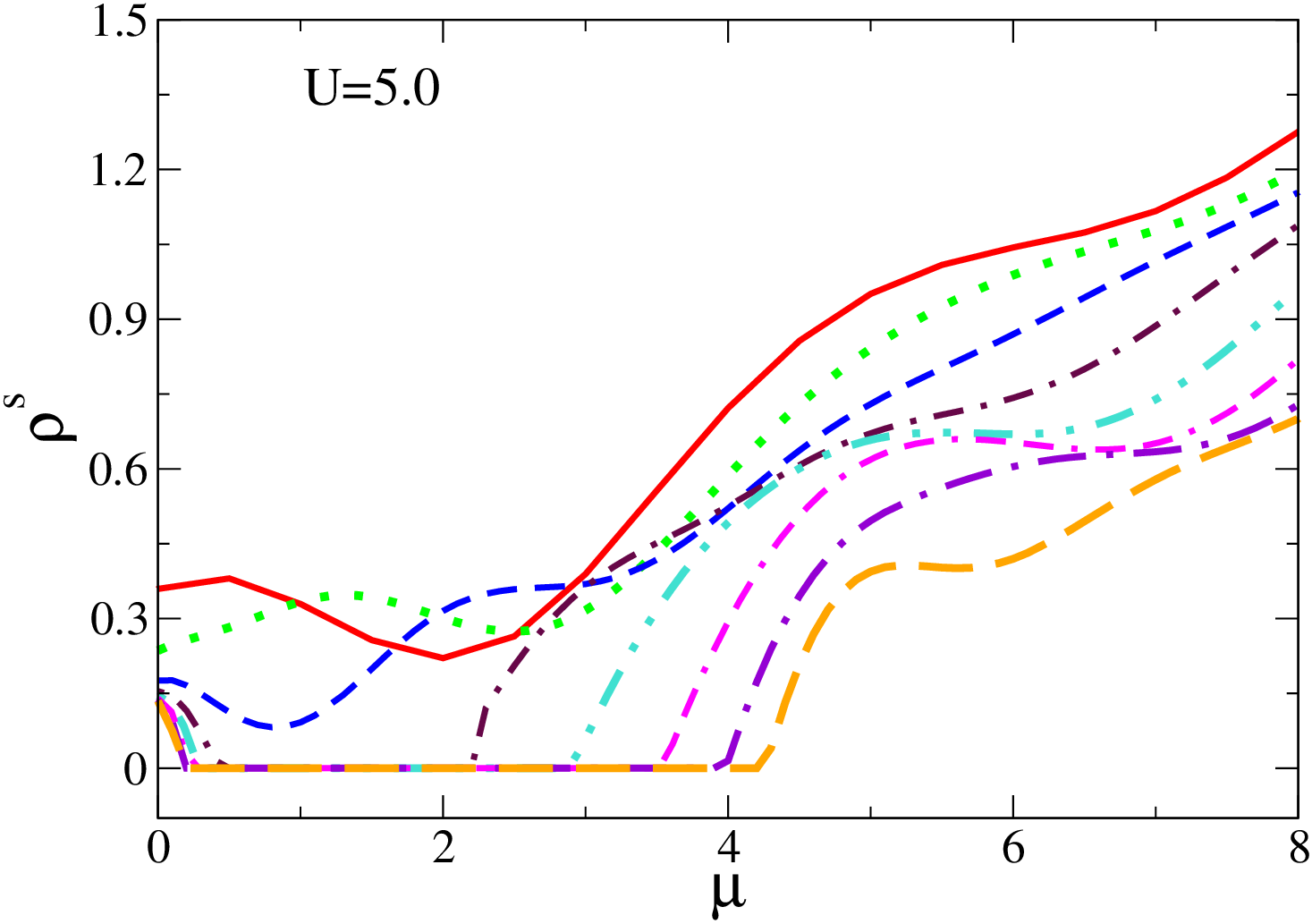,height=2.0in,width=2.5in,angle=0}
\caption{(Color online) Variation of average superfluid density of a
unit cell as a function of $\mu$ for the same set of parameters as
in Fig.~\ref{fig:fig3}.} \label{fig:fig4}
\end{figure}
\begin{table}[!b]
  \centering
  \caption{$U=2.0$}\label{tab:u2}
\begin{tabular}{|r|r|r|r|r|r|r|}
\hline
$\lambda$ &$\rho$ &$\mu$ &$\rho^s_{1}$ &$\rho^s_{2}$ &$n_{1}$ &$n_{2}$ \\
\hline \hline
0.5 &0.620 &0.00 &0.58 &0.45 &0.73 &0.51 \\
0.5 &1.022 &0.70 &0.87 &0.75 &1.12 &0.92 \\
0.5 &2.017 &2.60 &1.72 &1.57 &2.13 &1.91 \\
1.5 &0.490 &0.20 &0.52 &0.21 &0.77 &0.21 \\
1.5 &1.020 &1.20 &0.98 &0.62 &1.32 &0.71 \\
1.5 &2.020 &3.10 &1.87 &1.41 &2.34 &1.69 \\
2.5 &0.490 &0.47 &0.39 &0.08 &0.91 &0.08 \\
2.5 &1.030 &1.71 &1.11 &0.49 &1.53 &0.53 \\
2.5 &2.030 &3.61 &2.02 &1.26 &2.57 &1.49 \\
3.5 &0.500 &0.70 &0.07 &0.01 &0.99 &0.01 \\
3.5 &1.020 &2.10 &1.14 &0.33 &1.70 &0.34 \\
3.5 &2.030 &4.10 &2.16 &1.10 &2.79 &1.28 \\
4.5 &0.500 &0.40 &0.00 &0.00 &1.00 &0.00 \\
4.5 &1.000 &2.40 &1.08 &0.19 &1.82 &0.19 \\
4.5 &2.000 &4.50 &2.24 &0.91 &2.96 &1.03 \\
5.5 &0.500 &0.30 &0.00 &0.00 &1.00 &0.00 \\
5.5 &1.000 &2.60 &0.89 &0.10 &1.91 &0.10 \\
5.5 &2.000 &5.00 &2.36 &0.77 &3.18 &0.84 \\
\hline
\end{tabular}
\end{table}
\begin{figure}[!t]
 \centering
\psfig{file=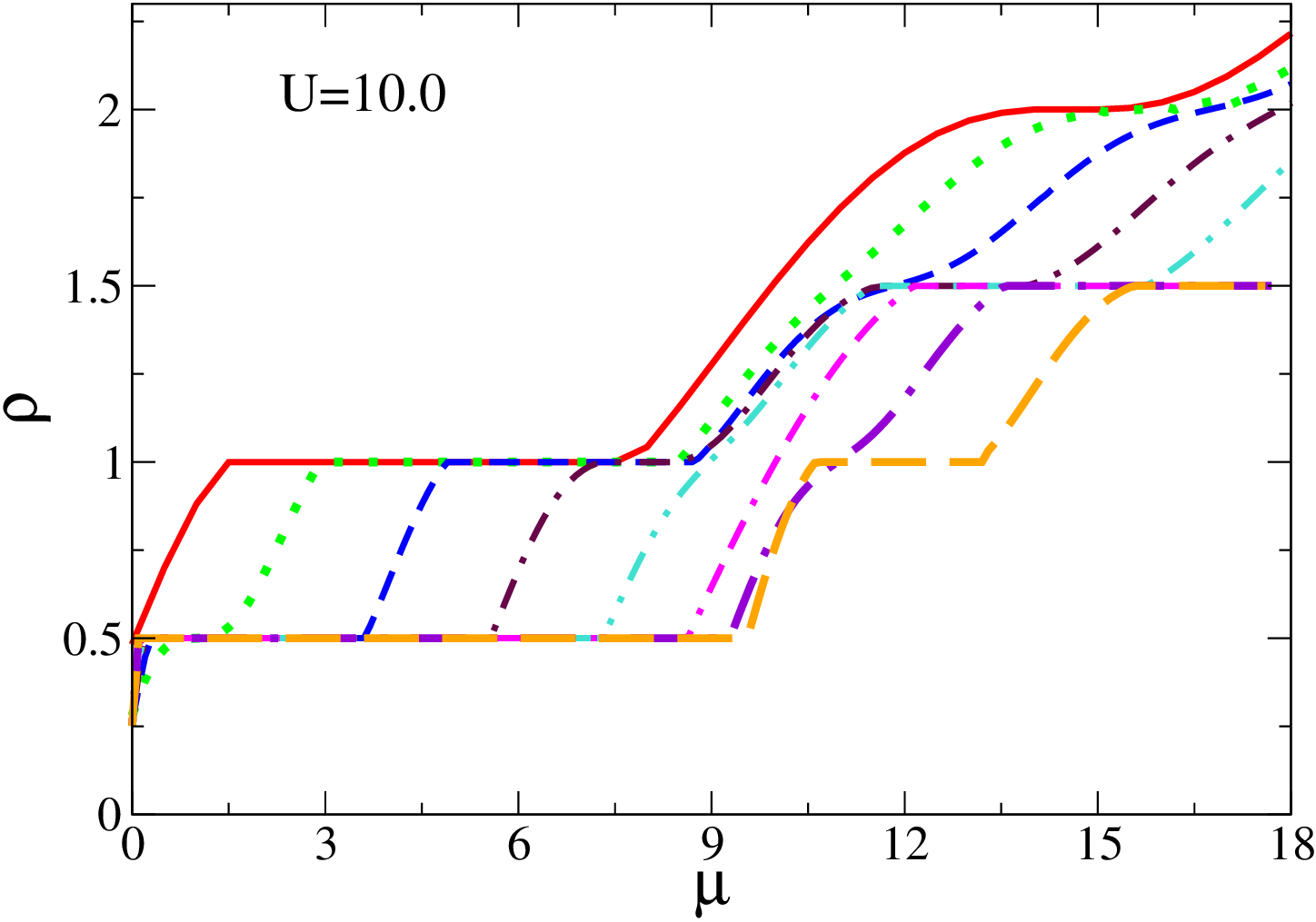,height=2.0in,width=2.5in,angle=0}
\caption{(Color online) Variation of average density of a unit cell
$\rho$ as a function of the chemical potential $\mu$ for U=10, but
for different values of $\lambda$, varying from 0.2 (red solid curve) to
14.2 (orange large dashed curve) at intervals of 2.0} \label{fig:fig5}
\end{figure}
This insulator phase is different from the standard Mott insulator
phase arising due to the on-site interaction. Here $U=2$ is small
and the Mott insulator phase is not expected. The reason for the
formation of an insulator phase for $U=2$ is due to the superlattice
potential, and to distinguish this insulator from the Mott insulator
phase, we call it as superlattice induced Mott insulator
(SLMI)~\cite{arya} as mentioned earlier. In order to understand the
SLMI phase, the distribution of bosons within the unit cell is
tabulated in the Table~\ref{tab:u2}. As we discussed in the previous
section, the unit cell consists of  $2^d$ sites and each cell has
two distinct sites, which we refer to as 1 and 2. The values of site
densities $\rho_1$, $\rho_2$ and superfluid densities $\rho^s_1$,
$\rho^s_2$ are listed in the table for different values of
$\lambda$. For $\lambda < 3.7$, the on-site superfluid densities
$\rho^s_1$ and $\rho^s_2$ remain finite for all densities. However,
for $\lambda \geq 3.7$ and density $\rho=1/2$, we find
$\rho^s_1=\rho^s_2=0$ and $\rho_1=1,~ \rho_2=0$. This implies,
within the unit cell, one site is occupied and the other site being
empty. Since this unit cell repeats to cover the entire lattice, it
has every alternate site occupied and the other is empty like a
charge density wave (CDW) phase which normally arises due to the
nearest neighbor interaction. However, it should be noted here in
this work that the CDW like density distribution is due to the
superlattice potential and there is no nearest neighbor interaction
involved. Since the distribution of bosons follow a pattern
$[1~~0~~1~~0~~1~~0~~\cdots]$ in all the $d$ directions of the
lattice, we call this phase as SLMI-I to distinguish it from SLMI-II
discussed below. The Table~\ref{tab:u2} also confirms that there is
no insulating phase for densities $\rho=1$ and $2$.

\begin{figure}[!t]
 \centering
\psfig{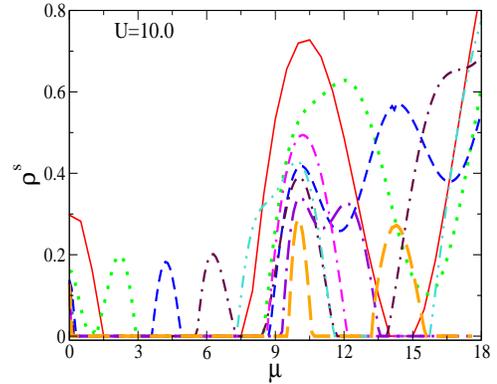}
\caption{(Color online) Variation of average superfluid density of a
unit cell as a function of $\mu$ for the same set of parameters as
in Fig.~\ref{fig:fig5}.} \label{fig:fig6}
\end{figure}
\begin{table}[!b]
  \centering
  \caption{$U=5.0$}\label{tab:u5}
\begin{tabular}{|r|r|r|r|r|r|r|}
 \hline
$\lambda$ &$\rho$ &$\mu$ &$\rho^s_{1}$ &$\rho^s_{2}$ &$n_{1}$ &$n_{2}$ \\
\hline \hline
0.2 &0.523 &0.0 &0.370  &0.35 &0.57  &0.48 \\
0.2 &1.030 &2.5 &0.270  &0.26 &1.03  &1.02 \\
0.2 &2.080 &7.5 &1.190  &1.17 &2.09  &2.06 \\
2.2 &0.500 &0.9 &0.110  &0.06 &0.95  &0.06 \\
2.2 &1.010 &3.3 &0.433  &0.36 &1.13  &0.90 \\
2.2 &2.000 &8.1 &1.250  &1.09 &2.19  &1.81 \\
3.2 &0.500 &0.5 &0.000  &0.00 &1.00  &0.00 \\
3.2 &1.000 &3.8 &0.570  &0.42 &1.22  &0.78 \\
3.2 &2.010 &8.7 &1.350  &1.08 &2.30  &1.73 \\
4.2 &0.500 &0.2 &0.000  &0.00 &1.00  &0.00 \\
4.2 &0.990 &5.3 &0.790  &0.29 &1.65  &0.34 \\
4.2 &2.010 &10.2 &1.540  &0.91 &2.62  &1.40 \\
7.2 &0.500 &0.2 &0.000  &0.00 &1.00  &0.00 \\
7.2 &1.010 &5.8 &0.620  &0.18 &1.82  &0.20 \\
7.2 &2.020 &10.7 &1.54  &0.83 &2.73  &1.30 \\
\hline
\end{tabular}
\end{table}

Results for $U=5$ are similar to those of $U=2$. In
Figs.~\ref{fig:fig3} and \ref{fig:fig4} we plot, respectively,
density $\rho$ and the superfluid density $\rho^s$ as a function of
$\mu$ for different values of $\lambda$. The system is in the
superfluid phase at $\rho=1/2$ and $\rho=1$ initially for low values
of $\lambda$ ($< 2.6$). For $\lambda > 2.6$, a plateau appears in
the $\rho$ versus $\mu$ plot for $\rho=1/2$ suggesting a gap in the
energy spectrum. The superfluid density vanishes in this region.
This plateau at $\rho=1/2$ widens as $\lambda$ increases. However,
the system remains in the SF phase at $\rho=1$ for all the values of
$\lambda$ considered. In Table~\ref{tab:u5}, we tabulate the values
of site densities and superfluid densities within the cell and we
conclude that the transition  from the SF to the SLMI-I phase is at
$\lambda=2.6$, when the superfluid density vanishes, and the
occupancy configuration is of the form [1 0 1 0 $\cdots$ ]. On the
other hand, at other values of $\rho$ and for all values of
$\lambda$, the superfluid densities, $\rho^s_{1}$ and $\rho^s_{2}$,
remain finite.
\begin{figure}[!t]
 \centering
\psfig{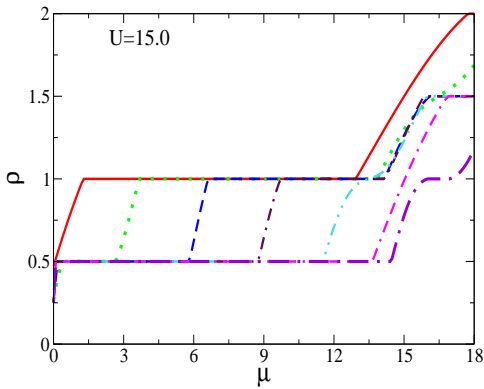}
\caption{(Color online) Variation of average density of a unit cell
$\rho$ as a function of the chemical potential $\mu$ for U=15, but
for different values of $\lambda$, varying from 0.2 (red solid curve) to
18.2 (violet large dot dashed curve) at intervals of 3.0} \label{fig:fig7}
\end{figure}

\begin{table}[!b]
  \centering
  \caption{$U=10.0$}\label{tab:u10}
\begin{tabular}{|r|r|r|r|r|r|r|}
 \hline
$\lambda$ &$\rho$ &$\mu$ &$\rho^s_{1}$ &$\rho^s_{2}$ &$n_{1}$ &$n_{2}$ \\
\hline \hline
0.2 &0.48 &0.00 &0.30  &0.29 &0.53  &0.44 \\
0.2 &1.00 &2.00 &0.00  &0.00 &1.00  &1.00 \\
0.2 &2.00 &14.50 &0.00  &0.00 &2.00  &2.00 \\
2.2 &0.50 &1.20 &0.00  &0.00 &1.00  &0.00 \\
2.2 &1.00 &3.00 &0.00  &0.00 &1.00  &1.00 \\
2.2 &2.00 &15.00 &0.00  &0.00 &2.00  &2.00 \\
6.2 &0.50 &0.50 &0.00  &0.00 &1.00  &0.00 \\
6.2 &1.00 &7.50 &0.00  &0.00 &1.00  &1.00 \\
6.2 &1.50 &12.00 &0.00  &0.00 &2.00  &1.00 \\
10.2 &0.50 &0.19 &0.00  &0.00 &1.00  &0.00 \\
10.2 &1.00 &9.97 &0.65  &0.32 &1.51  &0.47 \\
10.2 &1.50 &12.16 &0.00  &0.00 &2.00 &1.00 \\
14.2 &0.50 &0.10 &0.00  &0.00 &1.00  &0.00 \\
14.2 &1.00 &10.70 &0.00  &0.00 &2.00  &0.00 \\
14.2 &1.50 &15.60 &0.00  &0.00 &2.00  &1.00 \\
\hline
\end{tabular}
\end{table}
Results for $U=10$ are different from that of $U=5$. This difference
is mainly due to the fact that model~(\ref{eq:ham}) has SF to MI transitions
for integer densities. For $\rho=1$, the critical $U_c\sim 5.8$ for
the SF-MI transition~\cite{sheshadri} and this implies that for $U=10$, model
(\ref{eq:ham}) is in the Mott insulator phase for $\rho=1$. In the
Figs.~\ref{fig:fig5} and \ref{fig:fig6} we plot, respectively,
density $\rho$ and the superfluid density $\rho^s$ as a function of
$\mu$ for different values of $\lambda$. From these figures the
following conclusions are drawn. For small values of $\lambda$, the
plateau in the $\rho$ versus $\mu$ plot exist only for $\rho=1$ and
the $\rho^s$ vanishes in the same range of $\mu$ confirming the
expected SF to MI transition for $\rho=1$. The system remains in the
superfluid phase for all other densities. However, as we increase
$\lambda$, the plateau region at $\rho=1$ , i.e., the MI region, shrinks
first, completely disappears for some values of $\lambda$, and
re-appears again for higher values of $\lambda$.

A plateau develops at $\rho=1/2$ for $\lambda > 2.3$  and at
$\rho=3/2$ for $\lambda> 5.3$. From the tabulated values of $\rho_1$
and $\rho_2$ in the Table~\ref{tab:u10}, the insulator phase at
$\rho=1/2$ is the same as SLMI-I. The insulator phase at $\rho=3/2$
has a density distribution [2 1 2 1 $\cdots$ ] which we call
SLMI-III. The insulator phase at $\rho=1$ for higher values of
$\lambda$ has the occupation at alternate lattice sites [2 0 2 0
$\cdots$] which we refer to as the SLMI-II phase.

Thus for $U=10.0$, for $\lambda < 2.3$ the system exhibits a Mott
insulator phase for $\rho=1$ and SF phase else where. For $2.3 <
\lambda < 5.3$, the system has two insulating phases: SLMI-I for
$\rho=1/2$ and MI phase for $\rho=1$. The system is in the superfluid
phase for the rest of the densities. For $\lambda > 5.3$ the system
shows SLMI-I for $\rho=1/2$, MI phase for $\rho=1$, SLMI-III for
$\rho=3/2$ and SF for other densities. For $\rho=1$  the MI phase is lost
for $\lambda > 6.5$ and re-appears as SLMI-II for $\lambda > 13.1$.
The results for $U=10.0$ are in qualitative agreement with those obtained using DMRG~\cite{arya}.

\begin{figure}[!t]
 \centering
\psfig{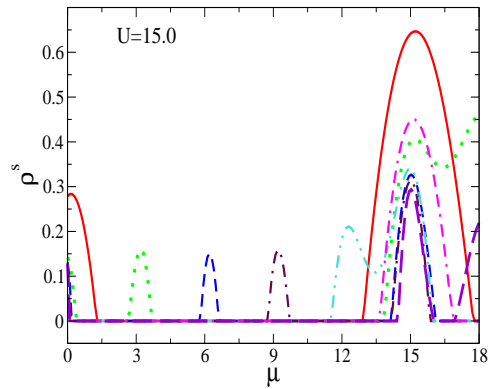}
\caption{(Color online) Variation of average superfluid density of a
unit cell as a function of $\mu$ for the same set of parameters as
in Fig.~\ref{fig:fig7}. } \label{fig:fig8}
\end{figure}

\begin{table}[!b]
  \centering
  \caption{U=15.0}\label{tab:u15}
\begin{tabular}{|r|r|r|r|r|r|r|}
 \hline
$\lambda$ &$\rho$ &$\mu$ &$\rho^s_{1}$ &$\rho^s_{2}$ &$n_{1}$ &$n_{2}$ \\
\hline \hline
0.2 &0.52 &0.10 &0.28  &0.28  &0.57  &0.47 \\
0.2 &1.00 &1.30 &0.00  &0.00 &1.00  &1.00 \\
0.2 &2.00 &17.80 &0.00  &0.00 &2.00 &2.00 \\
6.2 &0.50 &0.18 &0.00  &0.00 &1.00  &0.00 \\
6.2 &1.00 &6.66 &0.00  &0.00 &1.00  &1.00 \\
6.2 &1.50 &16.11 &0.00  &0.00 &2.00  &1.00 \\
15.2 &0.50 &0.10 &0.00  &0.00 &1.00  &0.00 \\
15.2 &1.02 &15.10 &0.60  &0.30 &1.55  &0.49 \\
18.2 &0.50 &0.09 &0.00  &0.00 &1.00  &0.00 \\
18.2 &1.0 &16.00 &0.00  &0.00 &2.00  &0.00 \\
18.2 &1.50 &19.08 &0.00  &0.00 &2.00  &1.00 \\
\hline
\end{tabular}
\end{table}

The system for $U=15.0$ behaves similarly as $U=10.0$. At
$\rho=1/2$, the  system starts off in the gapless SF phase for low
values of $\lambda(=0.2)$, as evident from the Figs.~\ref{fig:fig7}
and~\ref{fig:fig8} and Table~\ref{tab:u15}. But at $\rho=1.0$ and
$2.0$, the system is in the MI phase at this value of $\lambda$. As
$\lambda$ is increased to a value greater than $2.2$, a gap appears
at $\rho=1/2$, marking the transition from the SF to the SLMI-I
phase, as seen in the Table~\ref{tab:u15}, where we have vanishing
superfluid densities, and also a occupancy configuration of $[1~ 0~
1~ 0~ \cdots]$. Also, at $\rho=3/2$, another gap appears at
$\lambda=4.8$, implying the transition from the SF phase to the
gapped SLMI-III phase with a configuration $[2~ 1~ 2~ 1~ \cdots ]$.
As $\lambda$ becomes greater than $11.7$, the system at $\rho=1.0$
undergoes a phase transition from the MI to the SF phase, shown by
the non-zero values of the superfluid density. As $\lambda$ becomes
larger than $18.1$, the gap reappears once again, showing that the
system has entered into the gapped SLMI-II phase with configuration
$[2~ 0~ 2~ 0~ \cdots]$.

\section{Conclusions}

We have analyzed  the various phases exhibited by a system of bosons
in an optical superlattice with a unit cell consisting of two
distinct lattice sites using the mean-field decoupling
approximation, for various values of the superlattice potential,
$\lambda$, corresponding to four values of the on-site interaction
$U$. For $U=2.0$, we find that the system resides in the SF phase
for all densities for small values of $\lambda$. At $\rho=1/2$,
there is a transition from the SF to the SLMI-I phase at
$\lambda=3.7$, but for other densities, it remains in the gapless SF
phase. For $U=5.0$, the system undergoes a SF - SLMI-I phase
transition for $\rho=1/2$ at $\lambda=2.6$, but remains in the SF
phase for other densities and $\lambda$. For $U=10.0$,  the system
undergoes a SF - SLMI-I phase transition at $\lambda=2.3$ for
$\rho=1/2$,. However, for $\rho=1$, the system starts in the MI
phase, as the value of $U$ is large, and as $\lambda$ is increased,
the gap in the MI phase shrinks, and eventually goes to zero,
marking the MI-SF phase transition at $\lambda=6.5$. The system
stays in the gapless SF phase for $6.5 < \lambda < 13.1$. As
$\lambda$ is increased further the system undergoes a phase
transition from SF - SLMI-II at $\lambda=13.1$. For $\rho=3/2$, we
see a phase transition from SF to SLMI-III phase at $\lambda=5.3$.
Similar behavior is observed for $U=15.0$, with the system for
$\rho=1/2$ undergoing a phase transition at $\lambda=2.2$. For
$\rho=1$, the system makes a transition from the MI to the SLMI-I
phase at $\lambda=11.7$, and then from  the SF to the SLMI-II phase
at $\lambda=18.1$. Also a phase transition from the SF to the
SLMI-III phase at $\lambda=4.8$. It should be possible to extend
this calculation to superlattices with different periodicity. The
charge density wave order in the SLMI phase will depend on the
number of distinct sites within the unit cell. The mean field
approach is exact in the infinite dimension and the error, because
of neglecting the fluctuations, become severe in low
dimensions~\cite{rvpaispin1}. However, it proves to be an excellent
tool for the qualitative analysis (e.g. phase diagram), which is our
focus in this paper. Since the parameters of the Hamiltonian can be varied to a large range of values by tuning the 
strength of the optical potentials, we hope our detailed study of
model~(\ref{eq:ham}) will stimulate experimental studies that could
lead to the observation of the superlattice induced Mott insulators.

\section{Acknowledgement}
R.V.P. acknowledges financial support from CSIR and DST, India. We also acknowledge useful discussions
with Tapan Mishra and Gora Shlyapnikov.

\end{document}